\documentclass[aps,prc,preprint,superscriptaddress, amsfonts,amsmath, amssymb,
showpacs,floatfix]{revtex4}
\usepackage{graphicx,dcolumn}
\begin{document}
\title{Pair Correlations in Nuclei Involved in Neutrinoless Double Beta Decay: 
$^{76}$Ge and $^{76}$Se.}
\date{\today}
\author{S.J.~Freeman}
\affiliation{Schuster Laboratory, the University of Manchester, 
Manchester M13 9PL, UK}
\author{J.P.~Schiffer}
\email {schiffer@anl.gov}
\affiliation{Argonne National Laboratory,  Argonne,
Illinois 60439, USA}
\author{ A.C.C.~Villari}
\affiliation{GANIL (IN2P3/CNRS - DSM/CEA), B.P. 55027 14076 Caen Cedex 5, 
France}
\author{J.A. Clark}
\affiliation{A.W. Wright Nuclear Structure Laboratory, Yale University, New Haven, 
CT 06520, USA}
\author{C. Deibel}
\affiliation{A.W. Wright Nuclear Structure Laboratory, Yale University, New Haven, 
CT 06520, USA}
\author{S.~Gros}
\affiliation{Argonne National Laboratory,  Argonne,
Illinois 60439, USA}
\author{ A.~Heinz}
\affiliation{A.W. Wright Nuclear Structure Laboratory, Yale University, New Haven, 
CT 06520, USA}
\author{ D.~Hirata}
\affiliation{GANIL (IN2P3/CNRS - DSM/CEA), B.P. 55027 14076 Caen Cedex 5, 
France}
\affiliation{The Open University, Dept. of Physics and Astronomy, Milton Keynes, MK7 6AA, UK}
\author{ C.L.~Jiang}
\affiliation{Argonne National Laboratory,  Argonne,
Illinois 60439, USA}
\author{B.P.~Kay}
\affiliation{Schuster Laboratory, the University of Manchester, 
Manchester M13 9PL, UK}
\author{ A.~Parikh}
\affiliation{A.W. Wright Nuclear Structure Laboratory, Yale University, New Haven, 
CT 06520, USA}
\author{ P.D.~Parker}
 \affiliation{A.W. Wright Nuclear Structure Laboratory, Yale University, New Haven, 
 CT 06520, USA}
\author{ J.~Qian}
\affiliation{A.W. Wright Nuclear Structure Laboratory, Yale University, New Haven, 
CT 06520, USA}
\author{ K.E.~Rehm}
\affiliation{Argonne National Laboratory,  Argonne, Illinois 60439, USA}
\author{ X.D.~Tang}
\affiliation{Argonne National Laboratory,  Argonne, Illinois 60439, USA}
\author{ V.~Werner}
\affiliation{A.W. Wright Nuclear Structure Laboratory, Yale University, New Haven, 
CT 06520, USA}
\author{ C.~Wrede}
\affiliation{A.W. Wright Nuclear Structure Laboratory, Yale University, New Haven, 
CT 06520, USA}

\begin{abstract}

Precision measurements were carried out to test the similarities between the
ground states of $^{76}$Ge and $^{76}$Se.  The extent to which these two
nuclei can be characterized as consisting of correlated pairs of neutrons in
a BCS-like ground state was studied.  The pair removal (p,t) reaction was
measured at the far forward angle of 3$^\circ$. The relative cross sections
are consistent (at the 5\% level) with the description of these nuclei in
terms of a correlated pairing state outside the N=28 closed shells with no
pairing vibrations.  Data were also obtained for $^{74}$Ge and $^{78}$Se.
\end{abstract}
\maketitle


\section{introduction}

Interest in the possibility of observing neutrinoless double beta decay
(0$\nu$2$\beta$) is considerable.  If this decay were to be definitively
observed, it would show that the neutrinos are their own antiparticles.  In
addition, the rate of the decay would be a measure of the neutrino rest
mass, if the nuclear matrix element were known.  Unfortunately, theoretical
calculations of this do not agree well with each other \cite{Ba04}.  It
seems appropriate to determine additional properties of the ground states of
the possible 0$\nu$2$\beta$ systems by experiment, and thus help constrain
and test theoretical calculations of this exotic decay mode.  One of the
likely candidate nuclei is $^{76}$Ge decaying to the ground state of
$^{76}$Se.  We have started with a study of the properties of the ground
states of these nuclei, and especially the similarities and differences
between them, using transfer reactions.  One part of this study is an
accurate measurement of one-nucleon transfer in order to probe the
occupation numbers of valence orbits for both neutrons and protons, with
particular attention to changes in these occupations.  The other part is to
study pair-correlations in these nuclei by nucleon pair transfer.  Here we
report on a comparison of neutron pair transfer from the (p,t) reaction.  We
hope to obtain similar data on proton pair correlations from ($^3$He,n)
reactions in a future experiment.

Pair transfer between 0$^+$ states in the (p,t) or (t,p) reaction proceeds
via L=0 transfer and the angular distribution, for energies above the
Coulomb barrier, is sharply forward peaked. This feature was recognized
early \cite{Yo62} and was crucial in exploring the importance of such
correlations and their excitations, the so-called pairing vibrations
\cite{Be66}.  The latter are an indication of deviations from the simplest
pairing picture and can occur in regions of changing shapes, or when there
is a gap in single-particle states, such as near a shell closure.

We have carried out measurements of the neutron pair-removal (p,t) reaction
on targets of $^{74,76}$Ge and $^{76,78}$Se.  The reaction on $^{78}$Se was
measured because the $^{78}$Se(p,t)$^{76}$Se leads to $^{76}$Se, while 
$^{76}$Se(p,t)$^{74}$Se starts from the ground state of $^{76}$Se.  Thus
both reactions are relevant to the pairing structure of this ground state.
The $^{74}$Ge target was included as a check.

There are two relevant aspects to these measurements.  The first is the
matter of pairing vibrations.  The even Ge and Se isotopes are well studied
and evidence for excited 0$^+$ states has been established \cite{Gu77, Ar78,
Bo77} .  In some of the lighter isotopes of Ge and Se, two-neutron transfer
reactions have shown significant strength populating excited 0$^+$ states.
These pairing vibrations indicate that there are significant BCS-like pair
correlations in connecting the target ground state in an even initial
nucleus to excited 0$^+$ states in the final.  If there were significant
differences in this regard between reactions leading to or from $^{76}$Ge
and $^{76}$Se, this would be an indication that the pair correlations in the
initial and final states in double-beta decay differ. Any reliable
calculations of the process would presumably have to reproduce such
differences in order to obtain a reasonable matrix element for the
0$\nu$2$\beta$ decay. Secondly, if accurate cross sections were available
for pair transfer further checks could be made of the similarity between the
initial and final wave functions used in such calculations.


\section{Experimental Considerations}

Given that the L=0 (p,t) transitions are the strongest in the spectrum of
final states at very forward angles, which is also the region where the
approximations inherent in the distorted wave Born approximation (DWBA) are
best satisfied, it is desirable to carry out these measurements as close to
0$^\circ$ as feasible. The (p,t) reaction had been studied previously on Ge
\cite{Gu77} and Se \cite{Bo77} isotopes; relevant (t,p) measurements have
also been made \cite{Mo78, Le79, Wa87}. Both the experimental methodology
and the data analysis methods in the two (p,t) studies were different (the
target thickness was estimated from high-energy elastic scattering where
optical model predictions are ambiguous, the angular distributions start at
7.5 degrees, and only angle-integrated cross sections are quoted, etc.),
making a reliable systematic comparison of Ge and Se data difficult. Our
purpose here was to measure the cross sections at as far forward angles as
feasible and to obtain a consistent set of accurate cross sections with
particular care taken to reduce relative systematic uncertainties.

The choice of energy was governed by the desirability of having both protons
and tritons well above the Coulomb barrier.  For the present measurement we
therefore chose 23-MeV protons from the Yale ESTU tandem Van de Graaf
accelerator.  This energy is similar to that used in earlier experiments,
both for (p,t) and (t,p) reactions.  Because one of the objectives of the
present measurement was to obtain accurate cross sections, we chose to
measure the thickness of the evaporated germanium and selenium targets in
situ by simply lowering the proton beam energy to 6 MeV, where the elastic
scattering cross sections are very close to Rutherford values. The angle
should not be so far forward that small uncertainties in angle would become
significant and 30$^\circ$ was chosen because calculations with several
optical potentials showed that the deviation from Rutherford scattering was
less than 2\%.  The $^{76,74}$Ge and $^{78,76}$Se target thicknesses were
found to range between 160 and 400 $\mu$g/cm$^2$.  Since Se can sublimate at
a relatively low temperature, this low-energy target thickness measurement
was made at the beginning and then at the end of the experiment; no
significant differences ($<$3\%) were observed.  The highest beam currents
used were about 35 nA, though considerably lower (2-3 nA) for the 3$^\circ$
measurements.

The Yale Enge split-pole spectrograph was used for the measurements with a
focal-plane detector that cleanly separates tritons from other reaction
products. As monitors, two Si surface barrier detectors at $\pm$32$^\circ$
were used. They were calibrated in terms of beam intensity using a current
integrator connected to a Faraday cup.  The beam integrator was set to the
same scale that was used for the low-energy measurements and the solid-angle
setting for the aperture of the spectrograph was also the same, thus
establishing a relationship between an absolute cross section scale in terms
of the monitor yields, instead of the beam intergrator, for each target.

\begin{figure}
\includegraphics[width=13.0cm,angle=0]{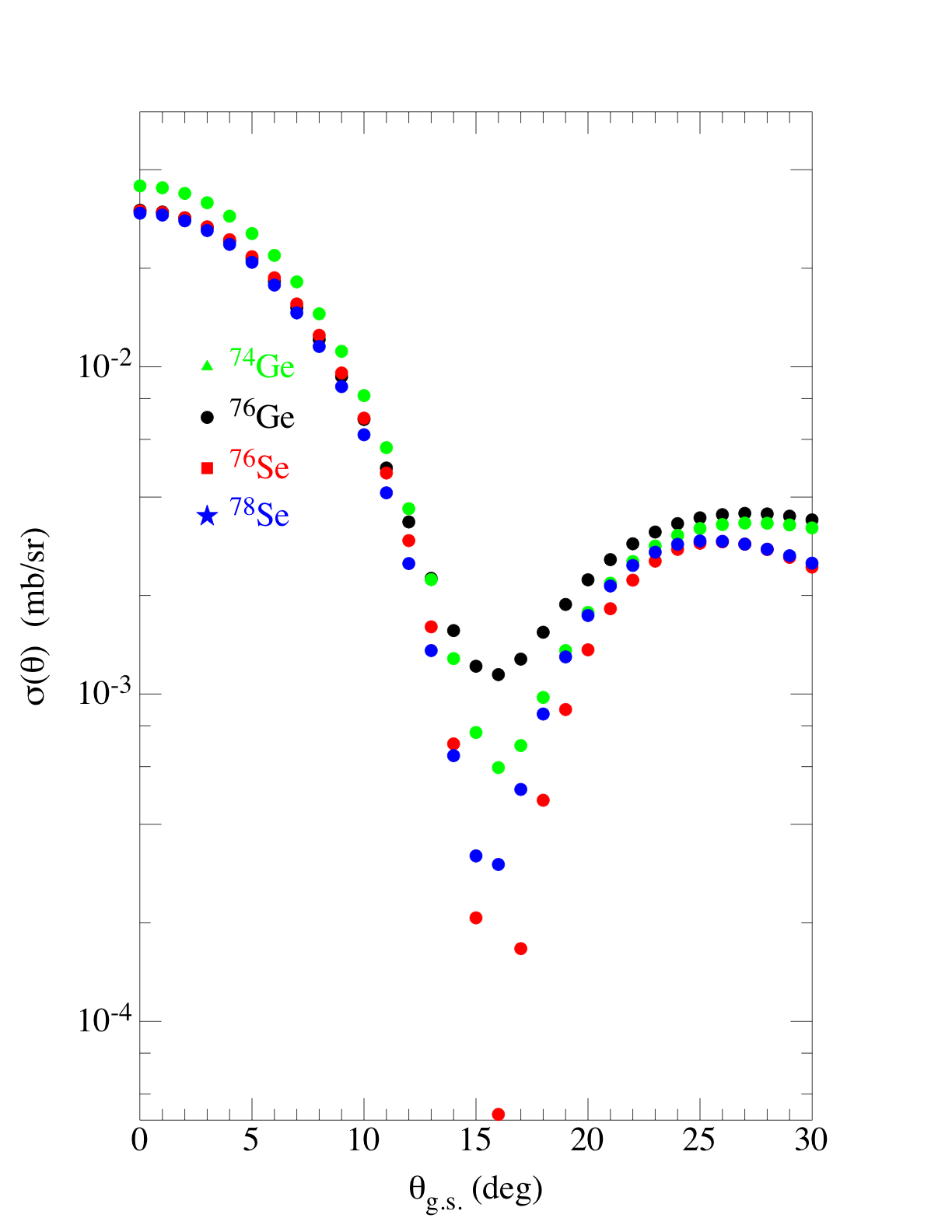}
\caption{\label{one}  (Color online) DWBA calculations of ground-state angular
distributions at 23 MeV for different targets, using the proton optical
potentials of Ref.~[12] and the triton potential from Ref.~[7].  The intent
of the figure is to illustrate the sensitivity of the shapes of the angular
distributions to the Q-values, while the peak cross sections remain
relatively stable.  The relative variation in the peak cross sections for
different choices of potentials is very similar. }
\end{figure}

The 3$^\circ$ setting for the spectrograph required that the Faraday cup be
retracted so that the beam-current measurement had to rely on the previously
calibrated monitor counters. Removal of the Faraday cup meant that the beam
entered the spectrometer. The magnetic rigidity of the tritons from the
reaction is such that, with the magnetic field set for observing tritons,
protons from the target cannot directly reach the focal plane and are
intercepted inside the spectrometer. Protons scattered at this point can
enter the focal plane and, whilst they can be distinguished from tritons by
their ionization density, they do impose a counting rate limit. As a result,
the furthest forward angle where measurements could be made was 3$^\circ$.
Distorted-wave calculations indicate that the cross section at this angle is
lower than that at 0$^\circ$ by about 8\%.  Spectra were also measured at
the laboratory angle of 22$^\circ$ , which is close to the minimum for L=0
angular distributions, though the location and depth of the minimum is very
sensitive to the Q-value and the distorting parameters as is seen in
Figure~\ref{one}.  Nevertheless, the ratio of the 3$^\circ$ to the
22$^\circ$ cross sections is huge compared to that for the other L values,
and is therefore an excellent identifier of L=0 transitions, though the
precise value of this ratio will depend on the exact location and depth of
the sharp minimum in the angular distribution.

\begin{figure}
\includegraphics[width=13.0cm,angle=270]{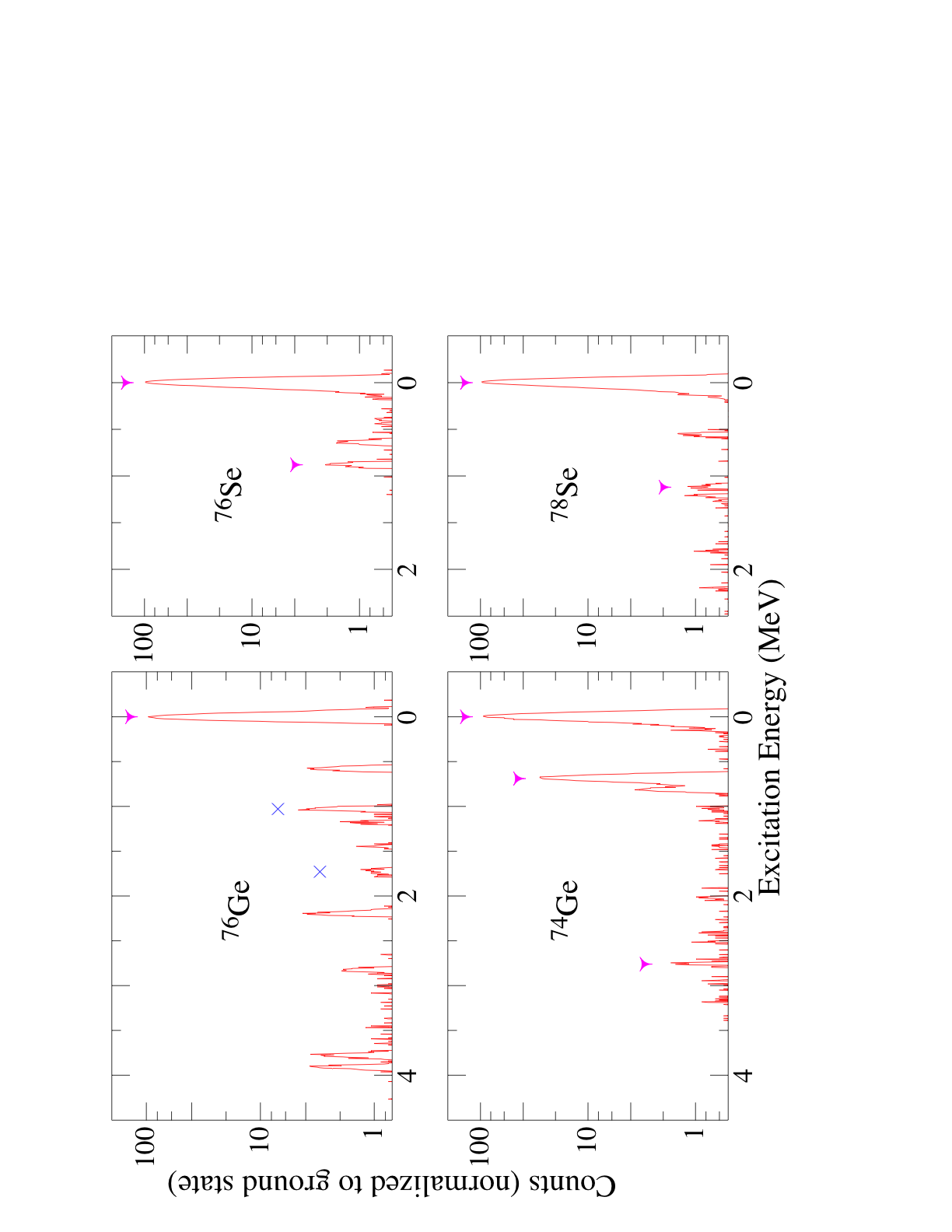}
\caption{\label{two} (Color online) Spectra of tritons at 3$^{\circ}$
measured with the Yale split-pole spectrograph, normalized to 100 for the
ground-state peak, and labeled in each case by the target nuclide. The peaks
corresponding to L=0 transitions are identified by a pointer.  Peaks due to
isotopic impurities are marked by an x.  While there is evidence in
$^{74}$Ge(p,t)$^{72}$Ge of substantial strength in a low-lying excited $0^+$
state, there are no large admixtures seen for the $^{76}$Ge and $^{76}$Se
targets.  }
\end{figure}

Representative spectra from the 3$^\circ$ measurements are shown in Figure~
\ref{two} where the ground-state transitions are clearly seen to dominate.
The results of the cross-section measurements are shown in Table~\ref{tone}
below, with ratios to the ground-state cross sections given only for states
with yields, at 3$^\circ$, larger than 1\% of the ground state yield. The
transitions where the ratio of cross sections between 3$^\circ$ and
22$^\circ$ is consistent with L=0 are shown in bold. More complete data,
though at further back angles and with less attention to accurate relative
cross sections had been reported, but the emphasis in these previous
measurements was on angular distributions starting at 7.5$^\circ$, and only
angle-integrated cross sections are quoted \cite{Gu77, Ar78}.  The
uncertainty in the present experimental cross sections is believed to be
$\pm$10\%, while that in the relative values is estimated at $\pm$5 \%.
These uncertainties are dominated by estimates of systematic errors
(constancy of the beam spot on target, accuracy of angle determinations in
the monitors, possible small drifts in monitor calibration, possible
inefficiency in the focal plane detector, uniformity of target thickness
etc.) while the statistical contribution is of the order of 1\%.

 The (p,t) differential cross sections at 3$^\circ$ for populating the 0$^+$
 ground states for the four targets are very similar: 6.4, 6.7, 6.0, and 7.1
 mb/sr for $^{74,76}$Ge and $^{76,78}$Se respectively.  Excited 0$^+$ states
 stand out in the ratio between the 3$^\circ$ and 22$^\circ$ yields, which
 is an order of magnitude larger than for any other excited state.  With the
 exception of the $^{74}$Ge target, none of these excited 0$^+$ states is
 populated with a cross section at 3$^\circ$ that is more than 2\% of that
 leading to the ground states.  In the $^{74}$Ge(p,t)$^{72}$Ge reaction the
 cross section to the first-excited 0$^+$ state is 1.9 mb/sr. This feature
 is well known \cite{Gu77} as an example of a pairing vibration. The case
 of $^{74}$Ge is illustrative of effects that can be problematic; however
 the context of the current work is related only to the $^{76}$Ge/$^{76}$Se
 double-beta-decay system.

 \section{DWBA Calculations}

The calculations were carried out with the DWBA program PTOLEMY \cite{PT} to
correct the dependence of the reaction on Q-values. The consideration of the
details of nuclear structure is beyond the scope of this study, even though
$^{76}$Ge and $^{78}$Se have 6 neutron vacancies in the N=50 shell,
$^{74}$Ge and $^{76}$Se have 8.  The form factor for the neutron pair was
calculated assuming a mass-2, $\ell=0$ di-neutron bound in a Woods-Saxon
potential with the appropriate binding energy and having 3 nodes in its wave
function.  The proton potentials were those of Ref.~\cite{Be69} and the
triton potential that of Ref.~\cite{Pe81}.  The measured cross sections at 3
degrees are given in Table~\ref{ttwo} together with the ratio of the
experimental cross sections to the calculated values. The absolute magnitude
of the DWBA cross section is very sensitive to the choice of the distorting
potentials as is the location of the first minimum in the angular
distribution. However, all calculations indicate that the relative values
between these targets of the calculated cross sections at 3$^\circ$ remain
the same (within 10\%) with various reasonable potentials for protons and
tritons.

\begin{figure}
\includegraphics[width=12.0cm,angle=0]{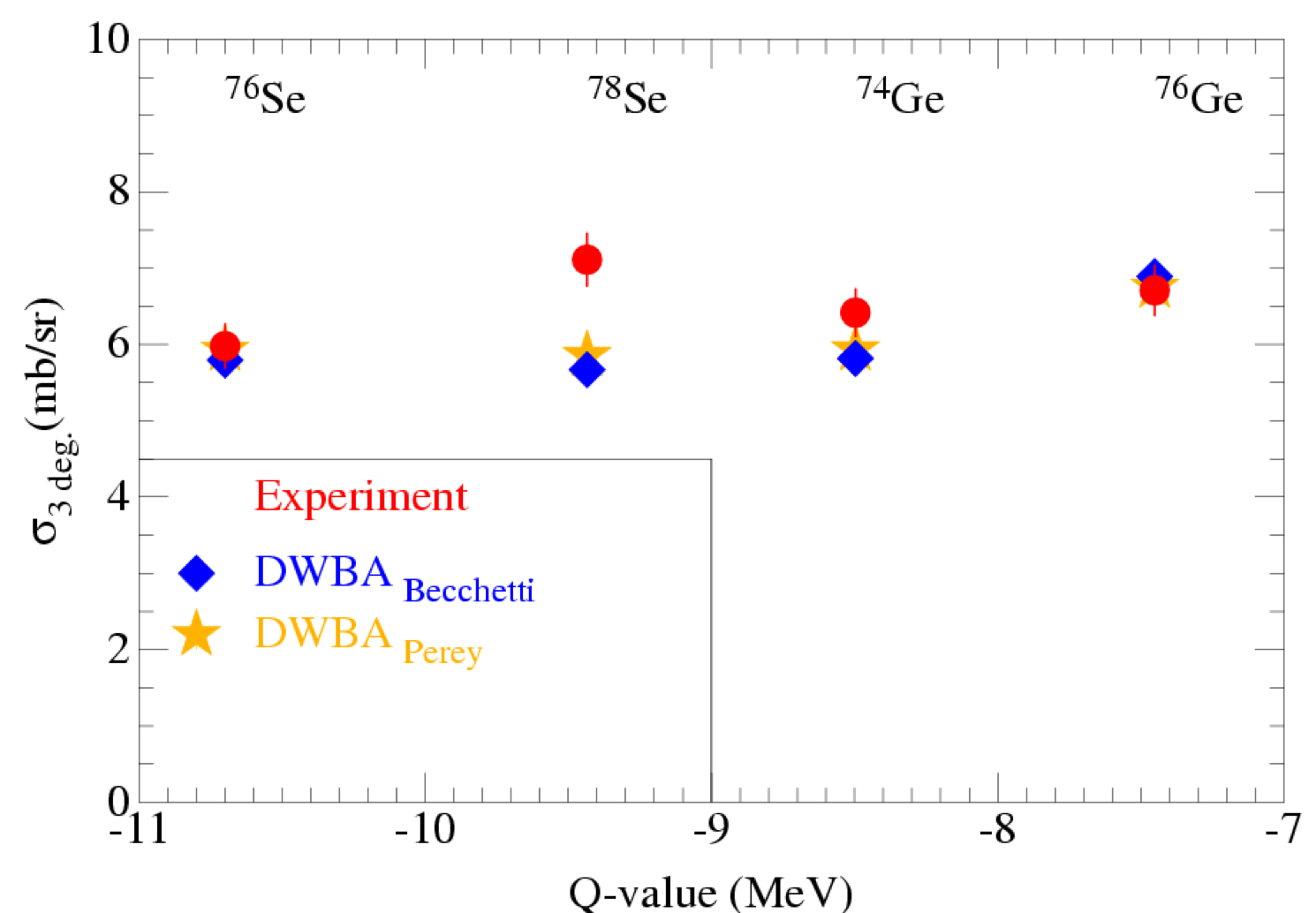} \caption{\label{three}
(Color online) The ground-state $0^+$ to $0^+$ cross sections at 3$^{\circ}$
are plotted as a function of Q-value, for convenience in display. Also shown
are the DWBA cross sections multiplied by one average normalization factor
for each proton potential. The estimated relative errors are shown on the
experimental points.  }
\end{figure}

 With the proton potential of Ref.~\cite{Be69}, the average ratio changes
from 136 to 217.  The experimental cross sections by themselves are
remarkably constant; without any correction for reaction dynamics they vary
by $\pm$6\%. Dividing the experimental numbers by the calculated reaction
cross sections, reduced the difference between $^{76}$Ge and $^{76}$Se even
further.  However, this improvement is probably not significant, in view of
the estimated 5\% relative error and the neglect of the different numbers of
neutrons. The peak cross sections and DWBA trends are also shown in
Figure~\ref{three}.

\section{Pair-adding reactions}

The pair-adding (t,p) reaction had been studied previously, for both the 
$^{74}$Ge(t,p)$^{76}$Ge \cite{Mo78,Le79} and the $^{76}$Se(t,p)$^{78}$Se
\cite{Wa87} reactions.  While the ground-state transitions are the same as
the ones studied here, transitions to excited states could, in principle,
show up and indicate pairing vibrations.  While the angular distributions
were not measured at as far forward angles as in the present work, some
estimate can be obtained by comparing integral cross sections.  No excited
0$^+$ states were seen in these (t,p) studies with a strength greater than
~4\% of the ground-state transition, confirming the dominance of pair
correlations in the ground states of these nuclei, with no splitting into
pairing vibrations.  

\section{Conclusions}

The experimental cross sections measured in this experiment are remarkably
constant for ground-state transitions with the various targets.  The
difference between $^{76}$Ge and $^{76}$Se, in the ratio of the experimental
ground-state transition strengths divided by the appropriate DWBA
calculations, is less than the 5\% estimated accuracy of the measurements.
Transitions to excited 0$^+$ states from $^{76}$Ge and $^{76}$Se targets are
no more than a few percent of the ground-state transitions indicating no
sign of the pairing vibrations that appear in some of the lighter isotopes.
Had such admixtures been present, this would have complicated a simple
comparison of the ground states.  The constancy of the ground-state strength
in pair correlations seems to be as true for neutron-pair-adding transfers
leading to these nuclei as it is for pair removal from them.  The present
results suggest that the ground states of $^{76}$Ge and $^{76}$Se exhibit
quantitatively very similar neutron pair-correlations.  Changes in pairing
are thus unlikely to be a significant complicating factor in the wave
functions of these states for calculations of neutrinoless double $\beta$
decay.

We wish to acknowledge helpful discussions with S.C. Pieper, Ben Bayman, and 
M.H. Macfarlane.  The help of John Greene in meticulous work on target
preparation is also acknowledged.  The work was supported by the U.S. 
Department of Energy, Office of Nuclear Physics, under contracts 
DE-FG02-91ER-40609 and DE-AC02-06CH11357, the UK Engineering and Physical 
Sciences Research Council, and IN2P3/CNRS-France.


\vfil\eject

\begingroup
\begin{table} 
\caption{\label{tone} Summary of (p,t) cross sections at 3$^\circ$ and 
the ratio (in \%) of these to the 22$^\circ$ values. Transitions consistent 
with L=0 are  shown in boldface.}
\begin{ruledtabular}
\begin{tabular}{ccc}
$^{74}$Ge(p,t)$^{72}$Ge&$\sigma_{\rm gs}(lab) = 6.4$~mb/sr\\
\colrule
Excitation  Energy (keV)&$\left(\sigma/\sigma_{\rm gs} \right)_{3^\circ}$
&Ratio(3$^\circ$/22$^\circ$)\\
\colrule
 \bf{ 0}&\bf{100}&\bf{86}\\
\bf{691}&\bf{29}&\bf{280}\\
834&2.8&0.9\\
1464&0.5&1.5\\
2024&0.5&4\\
\bf{2762}&\bf{0.9}&\bf{130}\\
\end{tabular}
\end{ruledtabular}
\vskip 2.5truemm
\begin{ruledtabular}
\begin{tabular}{ccc}
$^{76}$Ge(p,t)$^{74}$Ge    &  $\sigma_{\rm gs}(lab) = 6.7$~mb/sr\\
\colrule
Excitation  Energy (keV)&$\left(\sigma/\sigma_{\rm gs} \right)_{3^\circ}$&
Ratio(3$^\circ$/22$^\circ$)\\
\colrule
\bf{0}&\bf{100}&\bf{50}\\
596&3.2&1.0\\
1204&1.1&1.6\\
1463&2.2&0.8\\
2198&2.9&3\\
2833&1.7&6\\
\end{tabular}
\end{ruledtabular}
\vskip 2.5truemm
\begin{ruledtabular}
\begin{tabular}{ccc}
$^{76}$Se(p,t)$^{74}$Se&     $\sigma_{\rm gs}(lab)  = 6.0$~mb/sr\\
\colrule
Excitation  Energy (keV)&$\left( \sigma/\sigma_{\rm gs} \right)_{3^\circ}$&
Ratio(3$^\circ$/22$^\circ$)\\
\colrule     
\bf{0}&\bf{100}&\bf{115} \\
635&1.0&0.4\\
\bf{854}&\bf{1.4}&\bf{80}\\
\end{tabular}
\end{ruledtabular}
\vskip 2.5truemm
\begin{ruledtabular}
\begin{tabular}{ccc}
$^{78}$Se(p,t)$^{76}$Se&     $\sigma_{\rm gs}(lab)  =7.1$~mb/sr\\
\colrule
Excitation  Energy (keV)&$\left(\sigma/\sigma_{\rm gs} \right)_{3^\circ}$&
Ratio(3$^\circ$/22$^\circ$)\\
\colrule
\bf{0}&\bf{100}&\bf{150} \\
559&1.2&0.4\\
1121&0.8&4\\
1220&0.7&1.0\\
\end{tabular}
\end{ruledtabular}
\end{table}


\begin{table} 
\caption{\label{ttwo} 3$^\circ$ laboratory cross sections and ratios to
DWBA.  Cross sections are for the ground-state to ground-state transitions.}
\begin{ruledtabular}
\begin{tabular}{ccccc}
Target&$\sigma_{\rm exp}(lab)$ & $\sigma_{\rm DWBA}$ &
$\sigma_{\rm exp}$/$\sigma_{\rm DWBA}$\\
& (mb/sr) &  (mb/sr)& &\\

\colrule
$^{74}$Ge&6.4	&0.0438&147\\
$^{76}$Ge&6.7	&0.0499&135\\
$^{76}$Se&6.0	&0.0437&137\\
$^{78}$Se&7.1	&0.0431&164\\
\end{tabular}
\end{ruledtabular}
\end{table}

\endgroup

\end{document}